\documentclass[a4paper]{jpconf}
\usepackage{graphicx}
\begin{document}
\title{Moduli space volume of vortex 
and localization
}

\author{Akiko Miyake$^a$, Kazutoshi Ohta$^b$ and Norisuke Sakai$^c$}

\address{
$^a$Department of General Education, Kushiro National College of Technology, \\
Kushiro 084-0916, Japan, \\
$^b$Institute of Physics, Meiji Gakuin University, 
Yokohama 244-8539, Japan, \\
$^c$Department of Mathematics, Tokyo Woman's Christian University, 
Tokyo 167-8585, Japan
}

\ead{$^a$miyake@ippan.kushiro-ct.ac.jp; 
$^b$kohta@law.meijigakuin.ac.jp; \\
$^c$ Speaker, sakai@lab.twcu.ac.jp
}

\begin{abstract}
Volume of moduli space of BPS vortices on a 
compact genus $h$ Riemann surface $\Sigma_h$ is evaluated 
by means of topological field theory and localization 
technique. 
Vortex in Abelian gauge theory with a single charged 
scalar field (ANO vortex) is studied first and is found 
that the volume of the moduli space agrees with the previous 
results obtained more directly by integrating over the 
moduli space metric. 
Next we extend the evaluation to non-Abelian gauge groups 
and multi-flavors of scalar fields in the fundamental 
representation. 
We find that the result of localization can be consistently 
understood in terms of moduli matrix formalism wherever 
possible. 
More details are found in our paper\cite{Miyake:2011}. 
\end{abstract}

\section{Introduction}

Static solitons exert no force between them 
at the critical coupling, and are called BPS 
solitons\cite{Manton:2004tk}. 
Since the BPS solitons can coexist at arbitrary positions, 
the solutions of the BPS equations have many parameters 
such as the position of the soliton, which are called moduli. 
The moduli space of the BPS solitons plays important roles 
in understanding dynamics of solitons. 
For instance, the thermodynamical partition function 
can be obtained from the volume of the moduli space, 
since the solitons behave as free particles on their 
moduli space, when they move 
slowly\cite{Manton:1993tt,Shah:1993us}. 
Another non-trivial and important application of 
the volume of the moduli space is to obtain the 
nonperturbative effects, first found 
in the case of instantons. 
Nekrasov pointed out that the volume of the moduli 
space of the instantons, which is the so-called 
Nekrasov partition function, gives the non-perturbative 
effective prepotential of ${\cal N}=2$ supersymmetric 
Yang-Mills theory\cite{Nekrasov:2002qd}.  
To obtain finite values of the volume of the moduli space, 
we consider vortices on compact base manifolds 
such as genus $h$ Riemann surfaces. 
In recent years, BPS vortices in non-Abelian gauge theories 
have attracted much attention.\cite{Hanany:2003hp,Eto:2005yh,
Eto:2006mz,Eto:2006pg,Eto:2006uw,Eto:2007yv,Eto:2007aw,
Baptista:2008ex,Eto:2009wq,Manton:2010wu} 
The asymptotic metric of the moduli space is obtained for 
well-separated of non-Abelian vortices.\cite{Fujimori:2010fk} 
However, it is not enough to obtain the volume of the moduli 
space of the non-Abelian vortex. 

In another approach,  the ``localization'' 
technique\cite{Witten:1992xu,Blau:1993tv} 
of topological field theory has been 
applied successfully to evaluate the volume of the moduli 
space\cite{Moore:1997dj,Gerasimov:2006zt}. 
The K\"ahler structure of moduli space induces the 
localization property in the integration of the K\"ahler 
form which gives the volumes. 
The localization means that the integral of the 
volume form over the moduli space is localized 
(dominated) at isometry fixed points of the moduli space. 
This localization simplifies 
the evaluation of the volume of the 
moduli space drastically. 
The localization technique is based on a topological field 
theory, which can be understood as a twisted version 
of the supersymmetric theories\cite{Witten:1992xu}.

We apply the localization technique 
in the evaluation of the moduli space of the BPS 
vortices on Riemann surfaces in Abelian as well as 
non-Abelian $U(N_c)$ gauge theories. 
We consider $N_f$ flavors of Higgs fields in the 
fundamental representation extending the results 
in Ref.~\cite{Shah:1993us,Manton:1998kq} 
to the case of multiflavors, 
in the Abelian as well as the non-Abelian gauge theories. 
We find that our results by 
the localization technique completely agree with the 
previous results\cite{Shah:1993us,Manton:1998kq} 
for any topology of the Riemann surface 
for the ANO vortices ($N_c=N_f=1$). 
We can also evaluate the volume of the moduli space of the 
non-Abelian vortices, 
although it has been difficult to construct the metric 
of moduli space of the non-Abelian vortex apart from 
well-separated local vortices ($N_f=N_c$) \cite{Fujimori:2010fk}. 
By imposing the BPS equations as constraints in 
the field configuration space, we can regard the moduli 
space of BPS solitons as the quotient space of the fields, 
similarly to the K\"ahler quotient space. 
The integrals to give the volume of the moduli space 
reduces to simple residue integrals in the localization 
technique. 
Consequently we can evaluate the volume of the moduli 
space of the BPS vortices much easier than the explicit 
construction of the metric from the BPS solutions. 
We also work out the metric of the moduli space of 
single Abelian as well as non-Abelian vortices using 
 moduli matrix formalism\cite{Eto:2006pg} 
in order to compare it to 
our result from localization 
technique. 

\section{BPS vortices on Riemann surfaces}

Let us consider a $(2+1)$-dimensional space-time 
with the line element
\begin{eqnarray}
ds^2
= -dt^2+\sigma [(dx)^2+(dy)^2] 
= -dt^2+g_{z \bar z} dz d\bar z 
=g_{\mu\nu}dx^\mu dx^\nu, 
\label{eq:line_element} 
\end{eqnarray}
where the conformal factor and the complex coordinate 
are denoted as $\sigma=g_{z\bar z}$ and $z=x+iy$, 
respectively. 
We will denote the time coordinate $t$ and 
the spacial coordinates $x, y$ 
by $0$ and $i, j=1,2$, respectively, 
and space-time coordinates by 
$\mu, \nu =0, 1, 2$. 
We also define the K\"ahler 2-form 
$\omega=\frac{i}{2} g_{z\bar z}dz\wedge d\bar z$ from the metric. 
Then the area ${\cal A}$ of the Riemann surface 
$\Sigma_h$ is given by 
${\cal A}=\int dx dy \, \sigma =\int_{\Sigma_h}\omega$.

We are interested in a $U(N_{c})$ gauge theory in 
$(2+1)$-dimensional space-time with gauge fields 
$A_\mu$ as  $N_c \times N_c$ matrices 
and $N_{f}$ Higgs fields 
in the fundamental representation of the $SU(N_{c})_C$ 
$H$ as an $N_c \times N_f$ matrix. 
The Lagrangian is given in with the gauge coupling $g$ 
and the  Fayet-Iliopoulos (FI) parameter $c$ as 
\begin{eqnarray}
L&=&\int dx dy \sqrt{-{\rm det}(g_{\mu\nu})} \; \mathcal L
=\int dt dx dy  \; \sigma \; \mathcal L
\label{eq:unlagrangian1} 
\\
\mathcal L &=& \; {\rm Tr}\Bigl[- \frac{1}{2g^2} F_{\mu \nu}F^{\mu \nu} 
+ {\cal D}_\mu H ({\cal D}^\mu H)^\dagger 
-\frac{g^2}{4} \left( H H^\dagger - c{\bf 1}_{N_c} \right)^2 \Bigr],
\label{eq:uNLagrangianDensity}
\end{eqnarray}
where the covariant derivative and 
the field strength are defined by 
$D_{\mu} \equiv \partial_{\mu} -iA_{\mu}$ and 
$F_{\mu\nu} = i[D_\mu,D_{\nu}]$.  
This Lagrangian can be embedded into a supersymmetric 
theory with eight supercharges.

The Bogomol'nyi bound for the energy $E$ of vortices 
(depending on $x, y$ only) is obtained 
\begin{eqnarray}
E &=& 
\int dx dy \; {\rm Tr} \left[4D_{\bar z}H D_z H^\dagger 
+\frac{1}{g^2\sigma}
\left(F_{12}-\frac{g^2\sigma}{2} (c - H H^\dagger )\right)^2
+c F_{12}\right]
\nonumber \\
&\ge& c \int dx dy \; {\rm Tr} (F_{12})=2\pi c k, 
\label{eq:bogomolnyiBound}
\end{eqnarray}
where is the topological charge $k$ is the 
vorticity, namely the number of vortices. 
The bound is saturated if the following BPS equations 
\begin{eqnarray}
\mu_r &=& F - \frac{g^2}{2}(c-HH^\dagger)\omega,
\label{eq:unBPS1}\\
\mu_{\bar z} &=& {\cal D}_{\bar z} H, \quad 
\mu_z 
=
 {\cal D}_z H^\dagger,
\label{eq:unBPS3}
\end{eqnarray}
are satisfied. 
The Abelian gauge theory corresponds to $N_c=1$. 
Using these moment maps, the moduli space ${\cal M}_k$ 
of the vortex with the vorticity $k$ is given by a 
K\"ahler quotient space

\section{Abelian vortices}

Let us consider the volume of the moduli 
space of BPS vortices in Abelian ($G=U(1)$) gauge theory 
with $N_f$ Higgs fields on a compact Riemann surface 
$\Sigma_h$ of genus $h$.
The Higgs field $H(z,\bar z)$ has unit charge and is 
represented by an $N_f$ component vector.

We now introduce fermionic fields $\lambda$, $\psi$ 
to define BRST transformations for fields as follows
\begin{eqnarray}
\begin{array}{lcl}
QA = i\lambda, && Q \lambda = -d \Phi,\\
Q H = i\psi, && Q\psi = \Phi H,\quad
Q H^\dag = -i\psi^\dag, \qquad 
Q\psi^\dag = \Phi H^\dag,\\
QY = \Phi *\chi, && Q \chi = Y,\quad \quad 
Q \Phi = 0, 
\end{array}
\end{eqnarray}
where we have used form notations.  
These BRST transformations are nilpotent up to gauge transformations,
namely
$Q^2= -i\delta_{\Phi}$,
where $\delta_{\Phi}$ is the generator of the gauge transformation
with infinitesimal parameter $\Phi$.
Thus if we consider gauge invariant operators only, the 
BRST transformation $Q$ forms a cohomology for those operators, 
which is called the ``equivariant cohomology''. 
The equivariant cohomology clarifies topological aspects 
of (topological) field theory considering, 
and will play an essential but indirect role of the 
``localization'' in the evaluation of the volume.

$\Phi$ is BRST closed itself, so any function 
$
{\cal O}_0 \equiv {\cal W}(\Phi)
$ 
is also BRST closed. 
In the sense of the BRST symmetry, 
the 0-form operator becomes a good (physical) observable.
The 0-form observable satisfies the so-called descent 
relation 
\begin{eqnarray}
d{\cal O}_0 + Q {\cal O}_1=0,
\quad 
{\cal O}_1 \equiv \frac{\partial {\cal W}(\Phi)}{\partial \Phi}\lambda.
\end{eqnarray}
This fact means that the integral of ${\cal O}_1$ along a 
closed circle $\gamma$ on $\Sigma_h$ 
$\int_\gamma {\cal O}_1$ 
is BRST closed and a good cohomological observable. 
Similarly, ${\cal O}_1$ satisfies
\begin{eqnarray}
d{\cal O}_1 + Q {\cal O}_2=0,
\quad 
{\cal O}_2 \equiv i\frac{\partial {\cal W}(\Phi)}{\partial \Phi}F
+\frac{1}{2}\frac{\partial^2 {\cal W}(\Phi)}{\partial \Phi^2}
\lambda \wedge \lambda.
\label{eq:operator_descent2}
\end{eqnarray}
Thus the operator 
$\int_{\Sigma_h}{\cal O}_2 $ also becomes BRST closed. 
If we choose ${\cal W}(\Phi)=\frac{1}{2}\Phi^2$, we find 
that the integral 
$\int_{\Sigma_h}[i\Phi F+\frac{1}{2}\lambda\wedge \lambda]$ 
is BRST closed. Furthermore, we can see an integral 
$\int_{\Sigma_h}[i\Phi H H^\dag\omega +\psi^\dag\psi\omega]$ 
is also BRST closed, since the integrand can be written 
by the BRST exact form $Q[-iH\psi^\dag\omega]$. 
Therefore we obtain a BRST action 
\begin{eqnarray}
S_0= \int_{\Sigma_h}\left[
 i\Phi \mu_r + \frac{1}{2}\lambda\wedge\lambda
+\frac{g^2}{2}\psi^\dag\psi\omega
 \right]
\end{eqnarray}
as a BRST invariant completion of the BPS constraint 
(\ref{eq:unBPS1}), since 
$\Phi$ can be regarded as a Lagrange multiplier. 
We can implement the remaining BPS constraint (\ref{eq:unBPS3}) 
by introducing the BRST exact term $S_1$ to the action 
\begin{eqnarray}
S&=&S_0+S_1, 
\label{eq:action_volume}
\\
S_1&=& t_1 Q\int_{\Sigma_h}
i\chi \wedge * \mu_c,
\label{BRST exact action}
\end{eqnarray}
where $\mu_c \equiv \mu_z dz + \mu_{\bar z} d\bar z$. 
The volume of the moduli space of the BPS equations 
 (\ref{eq:unBPS1}) and (\ref{eq:unBPS3}) 
is obtained from the following integral 
\begin{eqnarray}
{\cal V}_k= \int {\cal D} \Phi {\cal D}^2 Y {\cal D}^2 \chi 
{\cal D}^2 A {\cal D}^2 
\lambda{\cal D}^2 H {\cal D}^2 \psi\, 
e^{-S}, 
\label{partition function of BPS eqs}
\end{eqnarray}
where the path integral is to satisfy the 
constraint $\frac{1}{2\pi}\int F=k$.

If the action includes BRST exact terms with couplings, 
the path integral does not depend on the couplings. 
Using this coupling independence of the  
integral, we can add the following BRST exact 
term to the action $S$ in Eq.~(\ref{eq:action_volume}) 
without changing the value of the integral 
\begin{eqnarray}
S_2 = t_2 Q\int_{\Sigma_h} i\chi \wedge * Y.
\end{eqnarray}
This part of the action serves to impose the remaining 
BPS constraint 
(\ref{eq:unBPS3}). 
By exploiting the coupling independence, we can go to a 
parameter region where the integral can be easily performed: 
let us take the limit $t_1\to 0$ and $t_2\to 1$ of the 
BRST exact terms. Then 
the action becomes 
\begin{eqnarray}
S' &=& \int_{\Sigma_h}\left[
 i\Phi \mu_r + \frac{1}{2}\lambda\wedge\lambda
+\frac{g^2}{2}\psi^\dag\psi\omega
 \right]
 + Q\int_{\Sigma_h} i\chi \wedge * Y\nonumber\\
 &=&  \int_{\Sigma_h}\bigg[
  i\Phi \left\{F-\frac{g^2}{2}(c-HH^\dag)\omega\right\} 
+ \frac{1}{2}\lambda\wedge\lambda 
+\frac{g^2}{2}\psi^\dag\psi\omega
+iY \wedge * Y + i\Phi \chi \wedge \chi
 \bigg]. 
\label{eq:finalBRSTaction}
\end{eqnarray}
Thus we can use the integral 
\begin{eqnarray}
{\cal V}_k=\int {\cal D} \Phi {\cal D}^2 Y  {\cal D}^2 \chi {\cal D}^2 A 
{\cal D}^2 \lambda{\cal D}^2 H {\cal D}^2 \psi\,
e^{-S'},
\end{eqnarray}
to evaluate the volume of moduli space ${\cal M}_k$.

First of all, we wish to integrate out the 
matter fields $H$, $\psi$, $Y$ and $\chi$, whose 
integrals are Gaussian. 
Neglecting the possible anomalies 
coming from the fermionic zero modes of matter fields, 
we obtain 
\begin{eqnarray}
{\cal V}_k= \int {\cal D} \Phi {\cal D}^2 A {\cal D}^2 \lambda \,
(i\Phi)^{N_f(\dim  \Omega^1\otimes {\cal L}_k 
- \dim  \Omega^0\otimes {\cal L}_k)}
e^{-\int_{\Sigma_h}\bigl[i\Phi(F-\frac{g^2c}{2}\omega)
+ \frac{1}{2}\lambda\wedge\lambda \bigr]},
\end{eqnarray}
where $\dim \Omega^n\otimes {\cal L}_k$ ($n=0,1$) stands 
for the number of holomorphic $n$-forms coupled with 
$U(1)$ gauge field (holomorphic line bundle) of the 
topological charge $k$. 
The Hirzebruch-Riemann-Roch theorem says 
\begin{eqnarray}
\dim  \Omega^0\otimes {\cal L}_k 
- \dim  \Omega^1\otimes {\cal L}_k 
= 1-h + \frac{1}{2\pi}\int_{\Sigma_h} F=1-h+k.
\end{eqnarray}
Thus we have 
\begin{eqnarray}
{\cal V}_k= \int {\cal D} \Phi {\cal D}^2 A {\cal D}^2 \lambda \,
\frac{1}{(i\Phi)^{N_f(1-h+k)}}
e^{-\int_{\Sigma_h}
\bigl[i\Phi(F-\frac{g^2c}{2}\omega)
+ \frac{1}{2}\lambda\wedge\lambda \bigr]}.
\label{eq:gauge_field_integral_volume}
\end{eqnarray}
By using 
$2(1-h)=\frac{1}{4\pi}\int_{\Sigma_h}R^{(2)}$ 
and $ k = \frac{1}{2\pi}\int_{\Sigma_h} F$ 
in terms of 
the curvature 2-form $R^{(2)}$ of the Riemann surface 
and the field strength 2-form $F$, 
we can exponentiate powers of $\Phi$ in 
Eq.~(\ref{eq:gauge_field_integral_volume}) to obtain 
\begin{eqnarray}
{\cal V}_k= \int {\cal D} \Phi {\cal D}^2 A {\cal D}^2 \lambda \,
e^{-S_{\rm eff}}, \quad 
S_{\rm eff} = S_R + S_F + S_V,
\end{eqnarray}
where
\begin{eqnarray}
S_R &=& \frac{1}{8\pi}\int_{\Sigma_h} 
\log (i\Phi) R^{(2)}, 
\quad
S_V =  -i\frac{g^2c}{2}\int_{\Sigma_h}
\Phi\omega
,\\
S_F &=& 
\int_{\Sigma_h} 
\left[
i\left( 
\Phi +\frac{1}{2\pi i} \log i\Phi \right)F 
+ \frac{1}{2}\lambda\wedge\lambda \right].
\end{eqnarray}
However $S_F$ is not invariant under the BRST symmetry 
(not BRST closed).
Since any regularization scheme should preserve the 
BRST symmetry, this means that we have  overlooked 
contributions from the fermionic zero modes in the 
integrals of fields $\psi, \chi$. 
To recover the contributions from the fermionic zero 
modes, we notice that the BRST closed action must 
take the form (\ref{eq:operator_descent2}) 
given by the descent relation 
\begin{eqnarray}
S_{F}' = \int_{\Sigma_h} \left[
i\frac{\partial {\cal W}_{\rm eff}}{\partial\Phi}F+\frac{1}{2}
\frac{\partial^2 {\cal W}_{\rm eff}}{\partial\Phi^2}
\lambda\wedge \lambda
\right], \quad 
\label{eq:additional_fermion}
\end{eqnarray}
with ${\cal W}_{\rm eff}(\Phi) = \frac{1}{2}\Phi^2 
+ \frac{N_f}{2\pi i}\Phi(\log i\Phi -1)$. 
Then we obtain 
\begin{eqnarray}
S_{F}' = 
\int_{\Sigma_h} 
\left[i\left( 
\Phi +\frac{1}{2\pi i} \log i\Phi \right)F 
+ \frac{\mu(\Phi)}{2}\lambda\wedge\lambda \right], \quad 
\mu(\Phi) = \frac{\partial^2 {\cal W}_{\rm eff}}{\partial\Phi^2} 
= 1+\frac{N_f}{2\pi i \Phi}.
\label{eq:mu}
\end{eqnarray}
The only correction due to (previously neglected) anomalies 
of the fermionic zero modes is changing the coefficient 
of $\frac{1}{2}\lambda\wedge\lambda$ from unity to $\mu(\Phi)$, 
which assures the BRST symmetry of the effective 
action.

Using $\int_{\Sigma_h}\omega ={\cal A}$ and $\int_{\Sigma_h} F = 2\pi k$, 
we integrate over $A$ and $\lambda$ and 
finally reduce the path integral into the 
following one-dimensional integral over 
ordinary real number $\phi$ (the 
constant mode of the field $\Phi$) 
\begin{eqnarray}
{\cal V}_k = \int_{-\infty}^{\infty} \frac{d\phi}{2\pi} 
\frac{\mu(\phi)^h}{(i\phi)^{N_f(1-h+k)}}e^{
i \phi \left( \frac{g^2c}{2}{\cal A} -2\pi k\right)}, 
\label{eq:zero_mode_integral1}
\end{eqnarray}
where $\mu(\phi)$ is defined in Eq.~(\ref{eq:mu}). 
Now let us evaluate the above integral. 
Since the integrand has a pole at $\phi=0$, we need to 
look for the correct integration contour to avoid the pole. 
The term $\int i\Phi HH^\dagger \omega$ in the action 
 (\ref{eq:finalBRSTaction}) of the path integral reveals 
that we need to choose the contour below the real axis 
in order to assure the convergence of path integral 
of matter fields $H$. 
Namely we should avoid the pole at $\phi=0$ 
counter-clock-wise below the pole. 
Expanding the integrand in powers 
of $\phi$, we can integrate 
term by term. We find that the volume is nonvanishing only if 
\begin{eqnarray}
\frac{g^2c}{2}{\cal A}-2\pi k \ge 0, 
\label{eq:bradlow}
\end{eqnarray}
 and 
\begin{eqnarray}
{\cal V}_k &=& 
\sum_{j=0}^{{\rm min}(h, d)} \frac{h!}{j!(h-j)!}
\left(\frac{N_f}{2\pi}\right)^{h-j}
\frac{1}{(d-j)!}\left( \frac{g^2c}{2}{\cal A}-2\pi k\right)^{d-j}
, 
\label{Abelian Nf}
\end{eqnarray}
where we have defined 
$d\equiv kN_f+ (1-h)(N_f-1)$ and used the residue 
integral formula. 
Eq.(\ref{eq:bradlow}) means that there is no solution 
for ${\cal A} < \frac{4\pi}{g^2 c} k$. 
This result is in agreement with the bound found in the 
case of ANO vortices ($N_c=N_f=1$), 
which is known as the Bradlow bound\cite{Bradlow:1990ir}. 
More interestingly, 
the non-vanishing volume exists 
only if 
$d\geq 0$, namely $k \geq (h-1)\frac{N_f-1}{N_f}$.
So we can choose any 
non-negative $k$ for $h=0,1$ 
or $N_f=1$, but $k$ must be sufficiently large 
for $h>1$ in the case of $N_f>1$.

Let us consider the simplest case $N_f=1$, namely the 
ANO vortices.  
In this case, 
Eq.~(\ref{Abelian Nf}) reduces to
\begin{eqnarray}
{\cal V}_k =
\sum_{j=0}^{{\rm min}(h, k)} 
\frac{ h!}{j!(k-j)!(h-j)!}
 \left(\frac{g^2c}{2}{\cal A}-  2\pi k \right)^{k-j}. 
\end{eqnarray}
To single out the net contribution from the $k$ vortex sector, 
we can mod out the contribution from the vacuum to define 
$\tilde{{\cal V}}_k \equiv {\cal V}_k/{\cal V}_0$. 
We find that our result agrees exactly with the previous 
result of the volume of the moduli space obtained from the 
moduli space metric by Manton and Nasir\cite{Manton:1998kq}, 
apart from an intrinsically ambiguous normlization constant 
to define the moduli space metric of a single vortex moduli 
space. 

 For semi-local vortices on sphere, 
we find the volume of the moduli space of vortices 
as 
\begin{eqnarray}
{\cal V}_k(S^2) =
\frac{1}{(kN_f+N_f-1)!}\left(\frac{g^2c}{2}{\cal A}-2\pi k\right)^{kN_f+N_f-1}.
\label{eq:u1Nfvolume}
\end{eqnarray}

\section{Non-Abelian Vortex}

Similarly to the Abelian case, the BRST transformations 
for non-Abelian case is given by 
\begin{eqnarray}
QA = i\lambda, && Q \lambda = -d_A \Phi,
\quad Q \Phi = 0,
\nonumber \\
Q H = i\psi, && Q\psi = \Phi H,\quad
Q H^\dag = -i\psi^\dag, \quad 
Q\psi^\dag = H^\dag \Phi,\\
QY_z = i\Phi\chi_z && Q \chi_z = Y_z,\quad
QY_{\bar z} = -i\chi_{\bar z}\Phi, 
\quad 
Q \chi_{\bar z} = Y_{\bar z}, 
\nonumber 
\end{eqnarray}
where $d_A\Phi\equiv d\Phi-i[A,\Phi]$.
The volume of non-Abelian vortices 
can be obtained by evaluating the following path integral 
\begin{eqnarray}
S&=&S_0+t_1 S_1 + t_2 S_2, 
\\
S_0 &=& \int_{\Sigma_h} {\rm Tr} \left[
i\Phi \left\{F-\frac{g^2}{2}(c-HH^\dag)\omega\right\}
+\frac{1}{2}\lambda \wedge \lambda 
+ \frac{g^2}{2} \psi^\dag\psi\omega
\right], 
\nonumber \\
S_1 &=& Q\int_{\Sigma_h} d^2 z {\rm Tr}  \left[
\frac{1}{2}g^{z{\bar z}}(\chi_z \mu_{\bar z} + \mu_z \chi_{\bar z})
\right],
\quad
S_2 = Q\int_{\Sigma_h} d^2 z  {\rm Tr}  \left[
\frac{1}{2}g^{z{\bar z}}(\chi_z Y_{\bar z} + Y_z \chi_{\bar z})
\right]. 
\nonumber
\end{eqnarray}
To evaluate the volume, it is easiest to choose the parameters 
$t_1=0, t_2=1$. We obtain 
\begin{eqnarray}
{\cal V}_k^{N_c,N_f}(\Sigma_h) = \int {\cal D} \Phi {\cal D}^2 A 
{\cal D}^2 \lambda {\cal D}^2 H {\cal D}^2 \psi {\cal D}^2 Y 
{\cal D}^2 \chi \, 
e^{-S_0-S_2}. 
\end{eqnarray}

To integrate out the fields, we choose a gauge which 
diagonalizes $\Phi$ as 
$\Phi={\rm diag}(\phi_1,\phi_2,\ldots,\phi_{N_c})$. 
After integrating out $H$, $\psi$, $Y$, $\chi$ 
and off-diagonal pieces of $A$ and $\lambda$ together with 
the associated ghosts, the integral reduces to the 
$U(1)^{N_c}$ gauge theory 
\begin{eqnarray}
{\cal V}_k^{N_c,N_f} & =&  
\int \prod_{a=1}^{N_c}  ({\cal D} \phi_a {\cal D}^2 A_a
{\cal D}^2 \lambda_a) \,
\frac{\prod_{a\neq b}(i\phi_a-i\phi_b)^{\dim \Omega^0\otimes 
{\cal L}_{k_a} \otimes {\cal L}_{k_b}^{-1} 
- \dim \Omega^1\otimes {\cal L}_{k_a}\otimes 
{\cal L}_{k_b}^{-1}}}
{\prod_{a=1}^{N_c}(i\phi_a)^{N_f(\dim \Omega^0\otimes 
{\cal L}_{k_a} - \dim \Omega^1\otimes {\cal L}_{k_a})}} 
\nonumber\\
&& \qquad
\times e^{-\sum_{a=1}^{N_c} \int_{\Sigma_h}\bigl[i\phi_a 
(F^{(a)}-\frac{g^2c}{2}\omega)
+\frac{1}{2} \lambda_a\wedge\lambda_a\bigr] },
\label{eq:volume_path_integral_gauge}
\end{eqnarray}
where the diagonal $a$-th $U(1)$ gauge field, field 
strength and gaugino are denoted as $A_a$, $F^{(a)}$ 
and $\lambda_a$, and $k_a$'s are diagonal $U(1)$ 
topological charges 
$\frac{1}{2\pi}\int F^{(a)} = k_a$, which satisfies 
the constraint of the total topological charge 
$k=\sum_{a=1}^{N_c} k_a$.

Similarly to the Abelian case, 
after using the Hirzebruch-Riemann-Roch theorem and 
supplementing the effective action to satisfy the BRST 
invariance, we can integrate over  $A_a$ 
and $\lambda_a$ to obtain the following finite dimensional 
integral of zero mode $\phi_a$ of $\Phi_a$ 
\begin{eqnarray}
{\cal V}_k^{N_c,N_f} 
&\!\!\!\!=&\!\!\!\! \sum_{\sum_a k_a = k}\! (-1)^\sigma \!
\int \prod_{a=1}^{N_c} \frac{d\phi_a}{2\pi}
\frac{\mu(\phi)^h\prod_{a < b}(i\phi_a-i\phi_b)^{2-2h}}
{\prod_{a=1}^{N_c}(i\phi_a)^{N_f(1-h+k_a)}}
e^{2 \pi i \sum_{a=1}^{N_c}\phi_a({\cal A}-k_a)},
\end{eqnarray}
where $\tilde {\cal A} =\frac{g^2c }{4\pi}{\cal A}$, 
$\mu(\phi) = \prod_{a=1}^{N_c}\left(1+\frac{1}{2\pi i} 
\frac{N_f}{\phi_a}\right)$, 
and $\sigma =\frac{1}{2}N_c(N_c-1)(1-h)-\sum_{a<b}(k_a-k_b)$.

We will consider the sphere topology for the Riemann 
surface. Let us first  examine the result for $N_c=2$.  
 For semi-local vortices ($N_f> N_c=2$), 
we find the asymptotic power of ${\cal A}$ for large area 
${\cal A} \to \infty$ as 
\begin{eqnarray}
{\cal V}_k^{2,N_f}(S^2) \propto \tilde{\cal A}^{kN_f+2(N_f-2)}. 
\label{eq:asymp_power_semi2}
\end{eqnarray}
In contrast, the asymptotic power is much smaller 
for local vortices ($N_f=N_c=2$) 
\begin{eqnarray}
{\cal V}_k^{2,2}(S^2) =  
\frac{2 (2\pi)^{2k} \tilde{\cal A}^k}{k!} 
+ {\cal O}(\tilde{\cal A}^{k-1}). 
\label{eq:asymp_power_local2}
\end{eqnarray}
Precise values of the volume of local vortices are given by 
\begin{eqnarray}
{\cal V}_0^{2,2}(S^2) 
&
=
&
 2, 
\nonumber \\
{\cal V}_1^{2,2}(S^2) %
&
=
& 
2 (2\pi)^2(\tilde {\cal A}-1), 
\nonumber\\
{\cal V}_2^{2,2}(S^2) 
&
=
&
  \frac{2 (2\pi)^4}{2!}
\left(\tilde {\cal A}^2-\frac{20}{6}\tilde {\cal A}
+\frac{17}{6}\right), 
\nonumber\\
{\cal V}_3^{2,2}(S^2) 
&
=
&
  \frac{2 (2\pi)^6}{3!}
\left(\tilde {\cal A}^3-7\tilde {\cal A}^2
+\frac{331}{20}\tilde {\cal A}-\frac{793}{60}\right), 
\nonumber\\
{\cal V}_4^{2,2}(S^2) 
&
=
&
 \frac{2 (2\pi)^8}{4!}
\left(\tilde {\cal A}^4-12\tilde {\cal A}^3
+\frac{818}{15}\tilde {\cal A}^2-\frac{2336}{21}\tilde {\cal A}
+\frac{18047}{210}\right). 
\end{eqnarray}

Let us next consider $N_c=3$ case. 
The asymptotic power of ${\cal A}$ 
for semi-local vortices ($N_f>N_c=3$) 
is given by 
\begin{eqnarray}
{\cal V}_k^{N_c=3,N_f}(S^2) \propto \tilde{\cal A}^{kN_f+3(N_f-3)}. 
\label{eq:asymp_power_semi3}
\end{eqnarray}
Combining the 
$N_c=2, 3$ results 
(\ref{eq:asymp_power_semi2}) and (\ref{eq:asymp_power_semi3})
of the asymptotic power of ${\cal A}$ 
for semi-local vortices ($N_f>N_c$), we conjecture the 
asymptotic power for generic semi-local vortices $N_f>N_c$ as  
\begin{eqnarray}
{\cal V}_k^{N_c,N_f}(S^2) \propto \tilde{\cal A}^{kN_f+N_c(N_f-N_c)}. 
\label{eq:semilocalAsymptotic}
\end{eqnarray}
Similarly to the $N_c=2$ case, we observe the reduction of the 
asymptotic power of ${\cal A}$ for local vortices ($N_f=N_c$) 
compared to the semi-local vortices 
\begin{eqnarray}
{\cal V}_k^{3,3}(S^2) = \frac{3!}{k!}
\left(\frac{(2\pi)^{3} \tilde{\cal A}}{2}\right)^k 
+ {\cal O}(\tilde{\cal A}^{k-1}). 
\label{eq:asymp_power_local3}
\end{eqnarray}
Combining the results 
(\ref{eq:asymp_power_local2}) and 
(\ref{eq:asymp_power_local3}) 
of $N_c=2, 3$ cases, we conjecture the 
asymptotic power of ${\cal A}$ of the 
volume of $k$ local vortices for the general values of $N_c=N_f$ as 
\begin{eqnarray}
{\cal V}_k^{N,N}(S^2) = \frac{N!}{k!}
\left(\frac{(2\pi)^{N} \tilde{\cal A}}{N-1}\right)^k 
+ {\cal O}(\tilde{\cal A}^{k-1}). 
\label{eq:asymptoticPowerUn}
\end{eqnarray}

 In the cse of the local Vortex $N_c=N_f=3$, 
\begin{eqnarray}
{\cal V}_0^{3,3}(S^2) &=& 3! ,\nonumber
\\
{\cal V}_1^{3,3}(S^2) &=& 3!  \times 
\frac{(2\pi)^3}{2}
(\tilde{\cal A}-1),\nonumber
\\
{\cal V}_2^{3,3}(S^2) &=& \frac{3!}{2!}  \times 
\frac{(2\pi)^6}{2^2}
\left(\tilde{\cal A}^2-\frac{46}{15}\tilde{\cal A}
+\frac{36}{15}\right),\nonumber
\\
{\cal V}_3^{3,3}(S^2) &=& \frac{3!}{3!}  \times 
\frac{(2\pi)^9}{2^3}
\left(\tilde{\cal A}^3-\frac{31}{5}\tilde{\cal A}^2
+\frac{3641}{280}\tilde{\cal A}
-\frac{23249}{2520}\right). 
\end{eqnarray}

We have also computed the case of torus topology for the 
base manifold. More detailed results can be found in 
Ref.~\cite{Miyake:2011}. 

\section{Effective Lagrangian of Vortices}

In order to understand the reduction of the asymptotic power 
of ${\cal A}$ of local vortices ($N_f=N_c$) 
compared to semi-local vortices ($N_f>N_c$), 
we study the effective Lagrangian 
by using the moduli matrix formalism\cite{Eto:2006pg}. 

 For arbitrary vortex number, we first work out the 
effective Lagrangian of Abelian semi-local vortices 
($N_f>N_c=1$) on the sphere, using the strong 
coupling $g^2\to\infty$. 
In the strong coupling limit, 
the solution $\Omega$ of the master equation is solved algebraically 
$\Omega = H_0H_0^\dagger/c$. 
Therefore the boundary condition for the vortex number $k$ 
is satisfied by requiring that at least one of components 
of moduli matrix to be a polynomial of order $k$, 
and all other components to be at most of order $k$ 
\begin{eqnarray}
H_0^{(k)}(z) =\sqrt{c}\left(\sum_{j=0}^{k}a^{(1)}_{j}z^j, \cdots, 
\sum_{j=0}^{k}a^{(N_f)}_{j}z^j\right), 
\label{eq:u1moduliMatrix_kvortex} 
\end{eqnarray}
where at least one of the coefficients of the $k$-th power 
is non-vanishing: $a^{(j)}_{k}\not=0$. 
Compared to the usual noncompact plane, we emphasize 
two new features of the vortex moduli on the 
compact Riemann surfaces which are realized in the moduli 
matrix (\ref{eq:u1moduliMatrix_kvortex}). 
Firstly we allow the leading power of $z$ to be in any 
components. 
If we use global $SU(N_f)$ rotations combined with the 
$V$-transformations, it is possible to 
place the leading power to be in a particular component, 
say in the first component. 
This form is the usual choice for the moduli matrix on 
noncompact flat plane \cite{Eto:2007yv}. 
These new $N_f-1$ moduli may be regarded as an orientation 
of the vacuum at infinity and are nonnormalizable on 
noncompact plane:  $(a_k^{(1)}, \cdots, a_k^{(N_f)})/a_k^{(1)}$ 
after dividing out by $V$-transformations. 
These $N_f-1$ extra complex moduli 
parameters are present even in the case of vacuum 
($k=0$) on compact Riemann surfaces. 
Secondly the additional $N_f-1$ ``size'' moduli are retained 
on compact Riemann surfaces, since they become normalizable 
and are dynamical variables in the effective Lagrangian. 
More specifically, the standard moduli matrix on noncompact 
plane contains up to only $(k-1)$-th power of $z$ except 
in the first component. 
The $N_f-1$ coefficients of these $(k-1)$-th power represent 
``size'' of vortices, are nonnormalizable, and have to be 
fixed by the boundary condition on noncompact plane. 
Both of the ``vacuum'' and the ``size'' moduli become 
normalizable and provide additional $2N_f-2$ complex moduli 
on compact Riemann surfaces. 
Let us emphasize that these new moduli should be fixed 
by boundary conditions and will not be a genuine 
moduli parameter of the solution once we take the limit of 
noncompact base manifold, such as a plane. 

The K\"ahler potential of the effective Lagrangian 
has been obtained\cite{Eto:2006uw}. 
We generalize the formula to the case of curved manifold 
such as Riemann surfaces, and 
insert the moduli matrix (\ref{eq:u1moduliMatrix_kvortex}) 
to obtain the K\"ahler 
potential on the sphere 
\begin{eqnarray}
K^{(k)}={\cal A} c \int_{S^2} dx dy \frac{1}{\pi(1+|z|^2)} 
\log\left(\sum_{i=1}^{N_f}
\left|\sum_{j=0}^{k}a_j^{(i)}z^j\right|^2\right)
.
\label{eq:kahlerPotential_kortex}
\end{eqnarray}
We find that the integral is convergent and is proportional to 
${\cal A}$, indicating that all the moduli parameters take 
values of order ${\cal A}$. 
So we find that each complex moduli gives a power of 
$\cal A$.
Since there are $kN_f+N_f-1$ complex moduli, we obtain 
the volume of the moduli space asymptotically ${\cal A}\to\infty$ 
to be proportional to the $kN_f+N_f-1$ power of ${\cal A}$ 
\begin{eqnarray}
\hat{\cal V}_{k}^{1, N_f}(S^2)
=
\left(Nc\pi{\cal A}\right)^{kN_f+N_f-1}
\times \int \prod_{i=1}^{N_f}\prod_{j=0}^{k}
d{a}_j^{(i)} 
\prod_{i'=1}^{N_f}\prod_{j'=0}^{k}
d{\bar a}_{j'}^{(i')} 
\frac{\partial}{\partial a_j^{(i)}} 
\frac{\partial}{\partial \bar a_{j'}^{(i')}} 
\left(\frac{K^{(k)}}{c\pi{\cal A}}\right), 
\label{eq:volMod_sphere_u1_kVortex}
\end{eqnarray}
where the coefficient of $(Nc\pi{\cal A})^{kN_f+N_f-1}$ 
is given by an integral representation for $K^{(k)}$. 
This asymptotic power agrees with the result 
 (\ref{eq:u1Nfvolume}) of the topological field 
theory. 

We have also worked out the effective Lagrangian on sphere 
for non-Abelian semi-local vortices ($N_f>N_c>1$) \cite{Miyake:2011} 
by using the moduli matrix approach\cite{Eto:2006pg}. 
Similarly to the Abelian semi-local vortices, we have 
found that the each complex moduli contributes to 
an asymptotic power of ${\cal A}$ at the strong 
coupling limit, in agreement with the result 
(\ref{eq:semilocalAsymptotic}) of 
the topological field theory. 

In the case of non-Abelian local vortices with 
$N_f=N_c=N$, we have worked out previously the metric of 
a single vortex on a plane in the moduli matrix 
formalism\cite{Fujimori:2010fk} . 
We found that only the position moduli can 
be of order $\sqrt{\cal A}$, whereas other orientational 
moduli consists of ${\bf C}P^{N-1}$ with the radius 
of order $1/g\sqrt{c}$. 
 Moduli space of multi-vortices $k >1$ are symmetric 
product of $k$ moduli spaces of each single vortex 
except for separations of order smaller than the vortex 
size $1/g\sqrt{c}$. 
These facts imply that the orientational moduli can 
only give a finite volume unrelated to ${\cal A}$, 
whereas the vortex position can be of order $\sqrt{\cal A}$. 
Therefore the volume of the moduli space for $k$ local 
non-Abelian vortices is proportional to ${\cal A}^{k}$, 
which agrees with our result (\ref{eq:asymptoticPowerUn}) 
of the topological field theory.

\section{Conclusion}
The volume of moduli space of vortices are computed 
for $U(N_c)$ gauge theory with $N_f$ Higgs fields in 
the fundamental representation, using the 
localization technique of topological field 
theory. 

Volume of moduli space of ANO vortices ($N_f=N_c=1$) for 
any vortex number $k$ and for any genus $h$ 
of Riemann surfaces is obtained 
and agrees with the previous direct calculation using 
the effective Lagrangian. 

Volume of moduli space is obtained both for 
Abelian semi-local vortices ($N_f>N_c=1$) and 
 non-Abelian vortices ($N_c>1$). 

We find that the  asymptotic power of area ${\cal A}$ 
for ${\cal A}\to \infty$ is 
${\cal A}^{kN_f+N_c(N_f-N_c)}$ for semi-local vortices 
($N_f>N_c$), and that the power reduces to 
${\cal A}^{k}$ for local vortices ($N_f=N_c$). 

Reduction of asymptotic power is understood by noticing 
that internal modes other than position do not extend 
over the whole area ${\cal A}$ for local vortices. 

 Localization technique should be powerful to obtain the volume of 
moduli space of other solitons\cite{Ohta2011}. 

\section*{Acknowledgements}
One of the authors (K.O.) would like to thank M.~Nitta, 
K.~Ohashi and Y.~Yoshida 
for useful discussions and comments.
One of the authors (N.S.) thanks Nick Manton, Norman Rink, 
Makoto Sakamoto, and David Tong for a useful discussion.
This work is supported in part by Grant-in-Aid for
Scientific Research from the Ministry of Education,
Culture, Sports, Science and Technology, Japan No.19740120 (K.O.),
No.21540279 (N.S.) and No.21244036 (N.S.), and by Japan Society
for the Promotion of Science (JSPS) and Academy of Sciences of
the Czech Republic (ASCR) under the Japan - Czech Republic
Research Cooperative Program (N.S.).

\end{document}